%% file: main.tex
\documentclass[letterpaper,twocolumn,10pt]{article}
\usepackage{usenix2019_v3}

\usepackage{tikz}
\usepackage{amsmath}
\usepackage{filecontents}
\usepackage{url}
\def\UrlBreaks{\do\/\do-}
\makeatletter
\g@addto@macro{\UrlBreaks}{\UrlOrds}
\g@addto@macro{\UrlBreaks}{%
	\do\/\do\d%
}
\usepackage{breakurl}
\usepackage{graphicx}
\usepackage{booktabs,siunitx}
\usepackage{listings}
\usepackage{multirow}

\usepackage{adjustbox}

\begin{document}

\title{A Survey of Access Control Misconfiguration Detection Techniques}

\author{
{\rm Bingyu Shen}\\
UC San Diego
}
\maketitle

\begin{abstract}
Access control mechanisms have been adopted in many real-world 
systems to control resource sharing for the principals in the system. 
An error in the access control policy (misconfiguration) can easily cause 
severe data leakage and system exploitation. Researchers have developed 
several methodologies to detect the access control misconfigurations 
through data mining, testing, and verification for various applications. 
This survey will study the line of works to 
detect access control misconfigurations and discuss some future 
research directions.

\end{abstract}

\input{intro}

\input{background}

\input{overview}
\input{mining}
\input{verify}
\input{testing}
\input{ccs}

\input{discussion}

\section{Conclusion}
\label{sec:conclusion}
In this survey, we discussed several representative works on access control 
misconfiguration detection. Based on the source of ground truth security goals, this 
work classifies the previous techniques into three categories: data mining, 
verification and testing approaches. We surveyed ten related works that fall into the 
three categories. By comparing the required input, target problem and technique 
limitations of the previous approaches, we provide a comprehensive overview of the 
access control misconfiguration techniques. We also discussed our own work to tackle 
the limitations in previous works. In the end, we proposed two future directions worthy 
of exploration, including providing sysadmins with more usable access control 
configurations and identifying cross-component misconfigurations

\bibliographystyle{plain}
\interlinepenalty=10000
\bibliography{references}

\end{document}

%% file: intro.tex
\section{Introduction}
\label{sec:intro}

Access control is the \textit{de facto} mechanism used in various computer systems to 
prevent unauthorized access. 
Due to the ever-increasing complexity of software systems and intricate nature of 
access control policy, access control misconfigurations happen frequently in the 
real-world 
scenarios and are a major cause of security incidents. In fact, access control 
misconfigurations are ranked the top ten web 
application security risks among all the security incidents in 2017~\cite{owasp}.
Even subtle errors in access control policies can also cause severe security 
consequences such 
as data theft and system compromise, as exemplified in Table \ref{tab:incidents_case}.
For example, a misconfigured server of a billion-dollar consulting services company 
exposed private customer information, certificates, 40,000 passwords, and other 
sensitive data to the public~\cite{news:2017}.

\begin{table}
	{
		\small
		\centering
		\setlength\tabcolsep{2pt}
		\resizebox{\linewidth}{!}{
			\begin{tabular}{lll}
				\toprule
				{\bf Time} & {\bf Incident} & {\bf Organization} \\
				\midrule
				2016.6 & 154 Million voters records leaked~\cite{news:2016} & L2 
				Political
				\\
				2018.2  & 119,000+ personal IDs 
				exposed~\cite{incident:fedex}               & FedEx \\ 
				2018.3  & 42,000 patients information 
				leaked~\cite{LongIsland:2018}          & Huntington hospital \\ 
				2018.4  & 63,551 patients records 
				breached~\cite{Middletown:2018}            & Middletown medical \\ 
				2019.1  & 24 million financial records 
				leaked~\cite{elastic:201901}              & Ascension \\
				2019.7  & 140 thousand SSNs and DoB 
				leaked~\cite{leak1}              & Capital One \\
				2019.9  & 20 million citizen records 
				exposed~\cite{elastic:2019}              & Elasticsearch \\
				\bottomrule
			\end{tabular}
		}
		\caption{{\bf {
					Recent publicly-reported security incidents
					caused by access control misconfigurations.}}
		}
		\label{tab:incidents_case}
	}
\end{table}

Considering the severe consequences of access control misconfigurations, many previous 
works have focused on defending against errors in access control misconfigurations in 
different development stages. 
Similar to software testing, the misconfigurations can also be classified into three 
types depending on the stages: fault, error and failure.
A fault in the configuration refers to the root cause of misconfiguration, such as 
syntactic errors, or misconfigured user groups.
An error in the configuration refers to the incorrect state of configuration goals 
which is the manifestation of a fault, such as users are granted too much access 
because of the wrong role assignment. 
A failure refers to the consequences of the misconfigurations, such as unauthorized 
access causing data leakage.
Some previous works have focused on the detection of attacks (i.e. abnormal 
access) with intrusion detection systems (IDS), based in the signature of network 
traffic (signature-based) or anomalies in the statistics of resource usage or 
parameters (anomaly-based)~\cite{liao2013intrusion}. 
However, these works may not be able to 
detect the unauthorized access if no patterns were found, and can not trace back to the 
configurations even though the root cause is misconfiguration.

In this survey, we focus on detecting the misconfiguration errors before they cause 
security failures. The proactive detection approaches will provide assurance for 
configurations before they are used in the production phase. 
The detection techniques will answer a fundamental question: whether the configuration 
result meet the intended \textit{security goals}. 
System administrators configure 
the access control configurations to achieve their security goals, but errors may be 
introduced to the configurations.
Therefore, another representation of the security goals must be obtained to 
check whether the configurations have errors.
We classify the techniques to obtain the 
security goals into three different categories. The first is the verification approach, 
the security goal is represented by formal specifications of security properties, 
written by developers or administrators. Second, the testing approach uses the test 
cases represent the desired security goals, which could be manually specified or 
automatically generated to achieve a comprehensive coverage.
The last approach is the data mining approach, where the security goals are inferred 
from 
correct security configurations. This survey aims to provide an overview of the three 
approaches and also discuss our work in tackling the limitations in the before 
mentioned approaches. We will also discuss some new opportunities to be explored in 
this area.

The rest of this survey is organized as follows. In Section \ref{sec:background} we 
discuss the unique challenges compared to other misconfigurations and some basic 
background of access control configurations. In Section \ref{sec:overview}, we discuss 
an overview of the detection approaches for access control misconfigurations. We will 
discuss the detailed techniques of data mining, verification and testing in Section 
\ref{sec:mining}, \ref{sec:verify}, and \ref{sec:test}. We will present our own work in 
Section \ref{sec:ccs}. Section \ref{sec:discuss} provides an summary of previous works 
and identifies several new directions in this area. Section \ref{sec:conclusion} 
concludes this survey.




%% file: background.tex
\section{Background}
\label{sec:background}
In this section, we discuss the difference between access control 
misconfiguration with other types of misconfigurations. Then we discuss the 
characteristics of access control models with real applications. Finally, we 
introduce a brief background of XACML which was commonly studied in previous 
works.

\subsection{Other Types of Misconfigurations}
Many previous works has been done to detect and diagnose the 
misconfigurations~\cite{yin2011empirical, 
attariyan2010automating,rabkin2011precomputing,xu2015hey,
	xu2013not,xu2016early}. There are mainly two types of works, 
the first type focus on misconfigurations that will lead to unavailable 
functionality 
~(e.g. system crash)
~\cite{attariyan2010automating,rabkin2011precomputing,xu2015hey,xu2013not,
xu2016early}, 
and the other type focus on the configurations that lead to poor 
performance (e.g. latency increase, throughput downgrade)~\cite{attariyan2012x, 
wang2018understanding}.
In order to detect such misconfigurations, some constraints need to be 
extracted 
from source code or systems with correct functionality or specified by domain 
experts. These constraints can be leveraged to detect aberrant system 
behaviors. 
However, access control misconfigurations are fundamentally different because 
of the purposes and manifestation.
The access control misconfigurations will not cause 
performance issues or functionality errors, instead misconfigurations will 
allow too much or too little access which violates the security goals or 
expectations.
For the configurations which grant too much access, the systems will behave 
normally and no one will 
complain about it, which will be left unnoticed until there is a data breach 
accident. For example, Capital One found the data breach of more than 100 
million people four months after the initial misconfiguration~\cite{leak1}.
The unique characteristics make the techniques in previous works hard to be 
leveraged directly to tackle the problem of access control misconfigurations.

\subsection{Access Control Models and Configuration Formats}

Access control mechanism fundamentally follows the basic access control matrix 
model~\cite{lampson1974protection}. However, different access control models 
are used to achieve the the desired access control goals in different systems. 
The formats of the configuration in each model are also customized without a 
uniform standard. This brings new challenges to detect access control 
misconfigurations in different systems.

First, different applications may use different access control models to 
achieve the security goal based on their needs. For example, firewalls use the 
simplest model, Access Control List (ACL), to achieve the security goal. The 
ACL only specifies subject, action and object. When the systems care about who 
can pass access control rights to another subject, Discretionary Access Control 
(DAC) and Mandatory Access Control (MAC) models are defined. The Unix file 
system applies DAC since all users can share their files to others by changing 
file permissions. However, SELinux applies MAC where only certain users with 
high privileges can change the access rights. MySQL uses Role-Based Access 
Control (RBAC) where it assigns roles to subjects and each role has a certain 
set of permissions. The Apache HTTPD web server uses Attribute-Based Access 
Control (ABAC) where the policy is a boolean expression to evaluate a set of 
attributes, e.g. POST and GET are two possible values for the attribute of HTTP 
method.

Second, even for the same access control model, the configuration format (e.g. 
syntactic and 
semantics) may be different. For example, ABAC are widely adopted among web 
servers, but each server defines their own configuration format. XACML  
is developed as a standard to represent ABAC or RBAC models, which was 
supported by Sun and Oracle systems. The heterogeneity of access control 
misconfigurations bring new challenges of 
validating the configurations in real systems across multiple components with 
different model and formats.
Previous works with the testing or verification method mainly focus on one 
model with XACML format, as discussed 
in \S\ref{sec:verify} and \S\ref{sec:test}. We also briefly discuss XACML here 
as background.

\subsection{XACML}
XACML (eXtensible Access Control Markup Language) \cite{standard2013extensible} 
is a specialized language to represent access control policies in XML format. 
Applications using XACML usually separate the workflow into enforcement (Policy 
Enforcement Point), decision (Policy Decision Point) and management (Policy 
Administration Point) into different parts for ease of implementation and 
maintenance.

\textbf{Components of XACML.}
Access control policies can be represented in XACML at three levels: policy set,
policy, rule. The policies at each level need to specify three or more elements 
to be a complete policy: \textit{subjects}, \textit{resources}, 
\textit{actions}, and \textit{environments} (optional). Each element has one or 
more attributes to be valid. Conditions can also be set at the rule 
level to match against real requests. XACML defines \textit{target} as a set of 
subjects, resources, actions and environments to be met for a given request. If 
the target matches with a rule, a decision of Permit or Deny  be returned. When 
there exist two or more rules contradicting each other, the decision will 
depend on the rule combining algorithm, such as using first applicable rule.

As a comprehensive tool to support complex access control needs, XACML 
implements a set of functions to support logic operations (e.g., and, or, 
etc.), regular expressions, high order functions (e.g. anyOf, all of, etc.), 
and many other operations which incorporate interactions between rules. This 
brings both flexibility and challenge for implementing correct policies and 
makes verifying real policies more difficult. 

\textbf{Adoption of XACML.} XACML standard is mainly used to implement ABAC 
models or RBAC models, with several implementations such as SunXACML, 
Axiomatics, AuthzForce.
However, it is not widely adopted nowadays for several 
reasons~\cite{xacmlisdead}.
First, XACML policies are complex and hard to maintain when used by 
practitioners. This may be because the ABAC model is hard to implement, or the 
complexity of XML language.
Second, XACML standard requires a centralized authorization point (PEP), which 
makes it hard scale to cloud or distributed systems. 
Third, there is little application support to switch to XACML standard, which 
takes huge efforts to refactoring many applications. 


%% file: overview.tex
\section{Overview}
\label{sec:overview}

This section provides an overview of the access control misconfiguration 
detection techniques. The three approaches, data mining, verification and 
testing, focus on different aspects of misconfigurations, with security 
constraints coming from different sources. The representative works surveyed in 
this paper are classified as shown in Table~\ref{tab:overview}.

\begin{table}[h]
	{
		\centering
		\setlength\tabcolsep{2pt}
		\resizebox{\linewidth}{!}{
			\begin{tabular}{llll}
				\toprule
				Techniques & Data Mining & Verification & Testing\\
				\midrule
				Related works &Bauer et al.'s work\cite{bauer2011detecting} & 
				Margrav~\cite{fisler2005verification}& Martin et al's 
			
				\\
				&Bazz~\cite{das2010baaz} & Mohawk\cite{jayaraman2011automatic}&	
				work\cite{martin2007automated, 
					martin2007fault} 
				
				\\
				&EnCore\cite{zhang2014encore} & Hu et al's 
				& Bertolino et al.'s \\
				&Shaikh et al.'s work\cite{shaikh2010inconsistency}& 
				work~\cite{hu2008enabling}& 
				work~\cite{bertolino2012automatic, 
					bertolino2014coverage}\\
				\bottomrule
			\end{tabular}
		}
		\caption{{\bf {Representative works of three different techniques to 
		detect access control misconfigurations}}
		}
		\label{tab:overview}
	}
\end{table}

Three approaches have made different trade offs between human efforts, 
comprehensiveness and accuracy in obtaining an accurate representation of 
security goals as mentioned in Section \ref{sec:intro}.
The data mining approaches have the least human effort with the assumption that 
most configurations or access control policies and results are correct, which 
is reasonable in most scenarios. However, since it relies on learning, there 
exist false positives and false negatives in the produced security 
representations.
In the verification approach, both the access control policy models and 
security specifications are provided by the developers or sysadmins, which 
requires great human efforts and deep domain knowledge of the system. This 
makes the approach less practical.
In the testing approach, the test cases represent the correct security goals. 
However, the quality of test cases matters most for the misconfiguration 
detection, i.e., if some test cases are missing then the misconfiguration will 
be left unnoticed. Manually designing comprehensive test cases is onerous and 
almost impossible, thus most works focus on automatically generating 
comprehensive test cases for access control policies. 

%% file: mining.tex
\section{Data Mining}
\label{sec:mining}
Data mining can be used to detect errors in the access control 
configurations by learning correct access configurations or access logs. This 
line of works has one assumption is that most accesses or configurations are 
correct and can be inferred to represent the sysadmins' security goals. 
Therefore, some data mining 
algorithms can learn from the mostly correct configurations and can be used to 
detect the anomalies inside the configurations. 
We discuss several relevant techniques to detect the existing works.
\begin{figure}[h]
	\includegraphics[width=\linewidth]{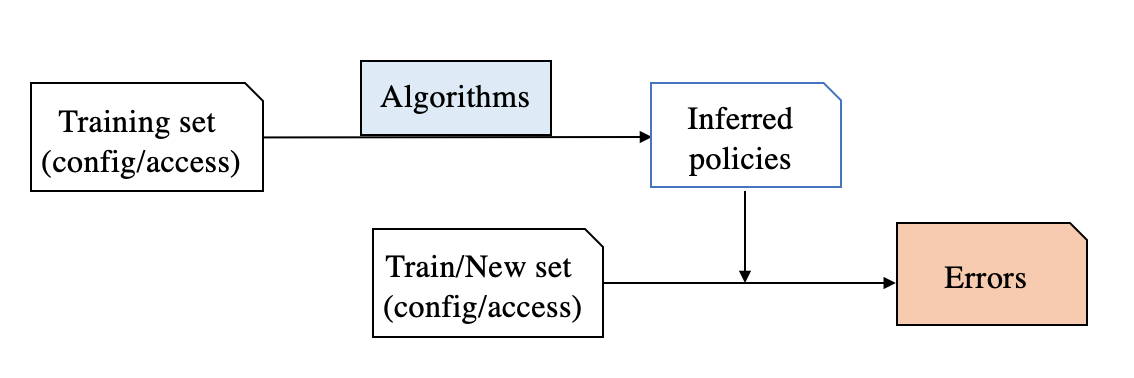}
	\caption{\textbf{Data mining approach overview.} The policies can be 
	inferred from configurations or access logs and be used on the original 
	training dataset or new set to detect 
	misconfigurations.\label{fig:datamining}} 
\end{figure}

\noindent \textbf{Bauer et al.'s work\cite{bauer2011detecting}} 
Bauer et al. focus on the problem that misconfigurations may falsely block 
legitimate accesses. They argue that this will cause user frustrations and 
waste administrators' time. Therefore, they applied\textit{ associate rule 
mining} to generate rules based on the access history logs, which are assumed 
to contain primarily correct access results. 
Then they use the generated rules to predict misconfigurations. The potential 
misconfigurations will be given to administrators for feedback - if the 
administrator is willing to fix the misconfiguration, then it is indeed a 
misconfiguration, otherwise not.

The intuition behind the associate rule mining (ARM) is that if several items 
appear together all the time, there exists some relationship between them. More 
formally, a rule $r$ can be represented as  
\begin{equation*}
S \in Dataset \implies T\in Dataset
\end{equation*}
which means if $S$ exists in the dataset, $T$ also exists in the dataset. The 
support metric refers to the frequency of a certain record. The 
confidence metric is used to measure the possibility of the rule to be 
true.
\begin{equation*}
\text{support}(r) = \frac{\text{total number of records contains both S and T}}
{\text{total number of records}} 
\end{equation*}
\begin{equation*}
\text{confidence}(r) = \frac{\text{total number of records contains both S and 
T}}
{\text{total number of records containts S }} 
\end{equation*}

They applied Apriori algorithm to mine the association 
rules~\cite{agrawal1994fast}. 
They first find all rules that have support more 
than a certain fraction of total records, which needs to be tuned in the 
algorithm. Second, they find all the rules whose confidence is higher than a 
certain level. 
The parameters are tuned in order to improve the quality of produced rules but 
avoid generating too many rules which do not have statistical significance. 

The developed rules have two uses. First, the rules can be used to detect 
whether misconfigurations exist in previous access logs. For example, if a 
rule's confidence level is very high, it means that some records where the 
premise holds but the conclusion does not hold are very suspicious.
Second, the rule can be used to audit new access logs. If the incoming records 
do not hold, they will be flagged and sent to the administrators.
To improve the prediction accuracy, they also involved the administrator 
feedback in the process to prune the produced rules.

Bauer et al's work have several limitations. First, the threshold parameter 
needs to be manually set to produce high-quality rules. The evaluation results 
show that low confidence and low support settings will improve the policy 
coverage, but the prediction accuracy will drop since too much noise is 
included in the produced rules.
Second, it assumes the access control configurations are static. The produced 
rules are based on all access logs. However, if the configuration changes, the 
rules will not be able to capture the changes in the history and produce false 
positives.
Third, the rules only handle ``accessibility'' problems where legitimate 
accesses are falsely denied. The other side of misconfigurations which may have 
security problems is not considered.
Fourth, even though the authors argue they can generate resolutions based on 
the detected misconfiguration, the misconfiguration actually refers only to the 
access result. Administrators still need to dig into the system to understand 
why the misconfiguration happens.

\begin{figure}
	\begin{minipage}{.6\linewidth}
		\centering
		\includegraphics[width=\linewidth]{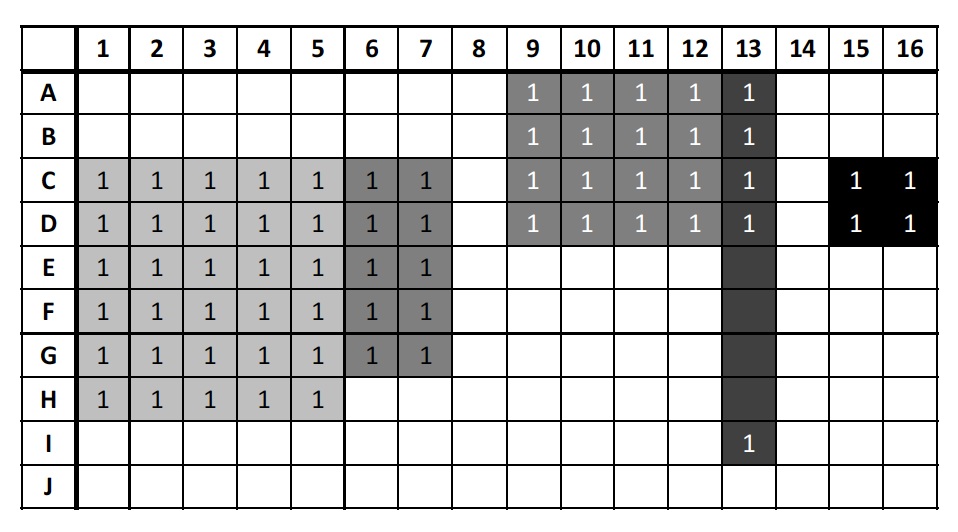}
		\caption{\textbf{Subject dataset~\cite{das2010baaz}} models users' 
		access rights to 
		resources, e.g.,  user A can access resource 9.}
		\label{fig:bazz1}
	\end{minipage}%
	\hspace{1em}
	\begin{minipage}{0.35\linewidth}
		\centering
		\includegraphics[width=0.6\linewidth]{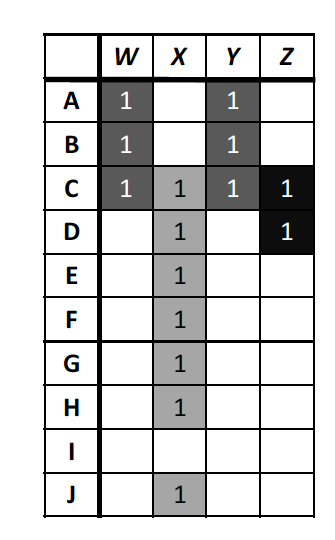}
		\caption{\textbf{Reference dataset~\cite{das2010baaz} }. e.g., user A 
		belongs 
		to group W 
		and Y. }
		\label{fig:bazz2}
	\end{minipage}
\end{figure}

\noindent \textbf{Bazz~\cite{das2010baaz}.} 
Bazz targets on the access control systems inside the organization with 
frequent access rights changes for individuals. The management of 
configurations is rather ad-hoc in the real world, which brings two possible 
problems. The first is that an individual does not have the right to certain 
resources which he/she should (accessibility problem). The second is that the 
individual is granted 
access which he/she should not have access (security problem). The first costs 
the administrators effort and time, while the second may 
bring real-world problems because of insider attacks. Bazz will be able to 
solve these two problems with anomaly detection techniques.

Bazz requires administrators to provide access control matrices at two levels 
as shown in Figure \ref{fig:bazz1} and \ref{fig:bazz2}. 
The first is the subject dataset, which defines users' access rights to resources 
with a matrix. The second is the reference dataset which defines users' 
relationships with groups in a matrix. 

Bazz takes three steps to detect the access control misconfigurations. First, 
Bazz generates summaries based on \textit{matrix reduction} of subject dataset 
and 
reference dataset. For example, in Figure \ref{fig:bazz1}, Bazz generates five 
summaries in different colors. One summary is $\{A,B,C.D\}\rightarrow 
\{9,10,11,12\}$, where the arrow means has access to. Note that the columns are 
disjoint in the summaries for later algorithm users.
Second, Bazz performs \textit{group mapping} to find outliers.
Bazz finds the outliers with two assumptions: (1) Users belong to the 
same group should have the same access rights to resources; (2) Users who do 
not belong to the same group should not have the same access rights.
Bazz also prioritizes the misconfiguration error reporting by the difference 
between unmatched users and total users in one group, since the assumption is 
that the majority will have correct permission settings.
Third, Bazz further performs \textit{object clustering} which only relies on 
the subject dataset. The intuition behind this algorithm is that by adding or 
removing some blocks in the matrix, the shape becomes more uniform. For 
example, if $H$ has access to 6 and 7, the rectangle becomes more uniform in 
Figure~\ref{fig:bazz1}. 

The main limitation of Bazz is that it requires domain knowledge from system 
administrators to specify the subject dataset and reference dataset. 
However, compared to completely switching to RBAC systems, the effort is much 
smaller. Also, Bazz designs monitoring stubs on clients to adapt to various 
kinds of subsystems, such as file server, git server, web servers. 
Administrators only need to specify the tasks to be monitored once. This helps  
detect misconfigurations in real-time. 
Another limitation is that the algorithms may detect false positives that 
wastes the administrator's time. However, the relatively low false positives may be 
worthwhile for detecting security misconfigurations which has disastrous 
consequences.

\begin{figure}
	\includegraphics[width=\linewidth]{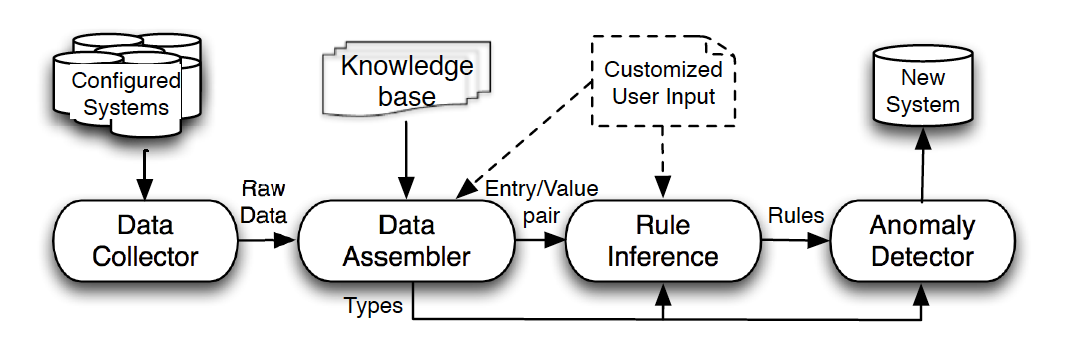}
	\caption{\textbf{The architecture of Encore.} Users can customize the rule 
	templates in the data assembler and rule aggregation metrics in the rule 
	inference part.~\cite{zhang2014encore}.\label{fig:encore}} 
\end{figure}

\noindent \textbf{EnCore~\cite{zhang2014encore}.} EnCore is an interesting 
framework to infer constraints from training 
configuration files. This approach also assumes that most configurations do not 
contain errors. With the inferred rules, EnCore applys the rules on other 
configuration files to detect violations. Even though EnCore targets on general 
configuration errors, such as data type mismatch or data value errors, it can 
also be applied to detect some permission errors in real systems, such as the 
owner of file directory mismatch. This is because EnCore provides users the 
ability to customize the rule templates in the rule inference phase.

The key idea of EnCore is to find correlated configuration entries and values. 
For example, \texttt{datadir=/var/lib/mysql} and \texttt{user=mysql} are two 
correlated configuration entries. EnCore further confirms this via system 
permission of \texttt{datadir} file path's ownership. Both the configuration 
file information and system environment information is collected in the 
initial data collection stage.

EnCore takes four steps to produce the anomaly detector as shown in 
Figure~\ref{fig:encore}. 
\begin{itemize}		
	\item First, EnCore collects all the data available in application 
	configurations as well as the system environment.
	\item  Second, with user rules, EnCore parses the raw data into uniform 
	key-value pairs, and infer the value types based on its syntactic and 
	semantic meanings. For example, if a string contains ``/'', it will be 
	inferred as file path.
	\item  Third, EnCore infers the rules based on rule templates the 
	configuration correlation. Similar to \cite{bauer2011detecting}, they first 
	apply association rule mining to replace the data types with concrete 
	values from configuration files. Then they will perform rule filtering 
	based on support and confidence levels. They also proposed to use entropy 
	to find configuration with frequent changes that are more likely to 
	useful. Users can also customize the rule templates in this stage.
	\item Fourth, with the inferred rules, EnCore inspects new configuration 
	files to detect rule violations.
\end{itemize}  

EnCore can be utilized to detect access control configuration errors, such as 
file system permissions, or whether IP subset configurations, via properly 
defined rule templates. The intuition is similar to Bauer et al's 
work~\cite{bauer2011detecting}, but EnCore focuses on mining at individual 
configuration entries level and predicts violations with produced common rules 
from correct configuration files.

Despite the benefits, EnCore also has several limitations. First, EnCore relies 
on administrators to write rule templates, which limits its scalability and 
comprehensiveness. Second, the simple key-value pairs of rules may not fit in 
the scenario that one configuration has multiple possible but correct enumerate 
values. Third, the interactions or logic operations between rules are ignored. 
For example, XACML has \texttt{anyOf} or ``first-applicable'' operators to 
combine the rules. This kind of more advanced rules can hardly be captured by 
rule templates.

\noindent \textbf{Shaikh et al.'s work~\cite{shaikh2010inconsistency}}
Shaikh et al. proposed to detect \textit{inconsistencies} in access control 
policies 
with decision tree algorithms. The inconsistencies in the access control 
policies may lead to abnormal behaviors or security vulnerabilities. The 
authors formally define the inconsistency as follows. Given a set of rules $R = 
\{R_1, R_2, ..., R_n\}$, a set of attributes $A=\{A_1, A_2, ..., A_n\}$, a 
decision category $C$ of Permit or Deny. Each role can be represented as 
$R_i:~~ A_1 \wedge  A_2\wedge ... \wedge A_n \rightarrow C$. Two rules are 
inconsistent if and only if their attributes are the same but the decision is 
different.
\begin{figure}
	\includegraphics[width=\linewidth]{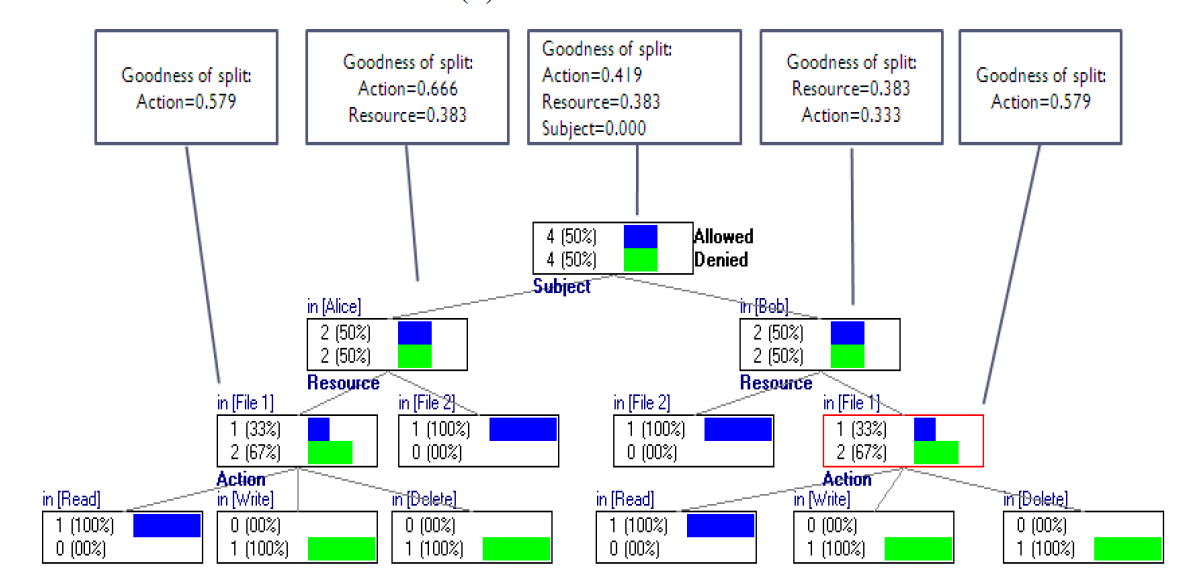}
	\caption{An sample decision tree with modified C4.5 algorithm
	~\cite{shaikh2010inconsistency}.} 
\end{figure}

The authors utilize C4.5 decision tree algorithm to train a 
classifier~\cite{safavian1991survey}. The decision tree will be split by 
attributes at each level. In each tree node, it contains policies with decisions 
of both Permit and Deny. An attribute can split the dataset S into P and D set.
Then the information gain can be defined as 
\begin{equation*}
\text{gain}(A) = I(S_P, S_D) - E(A)
\end{equation*},
where $I(S_P, S_D)$ is the total information to classify $S$ and $E(A)$ is the 
information needed to split into subtrees with attribute $A$. Suppose $A$ has 
$v$ possible values, there will be $v$ subtrees $S_1,...,S_v$.
\begin{equation*}
I(S_P, S_D) = -\sum_{i\in \{P, D\}}\frac{|S_i|}{|S_P| + 
|S_D|}log_2(\frac{|S_i|}{|S_P| + |S_D|})
\end{equation*}
\begin{equation*}
E(A) = \sum_{i=1}^{v} \frac{|S_{P_i}|+|S_{D_i}|}{|S_P| + |S_D|}  I(S_P, S_D)
\end{equation*}

The also modified the C4.5 algorithm in order to preserve as many attributes as 
possible in the decision tree. Each time they select the attribute with minimum 
information gain. In the terminal tree leaves, if there exists more than two 
decision categories, it will flag the node as an inconsistency. The original 
rule can be constructed from the leaf node to the root.

The main limitation of this work is that it only detects the inconsistencies in 
access control policies. 
However, the consistent access control policy may 
still have security flaws. 
Second, the tool requires deep domain knowledge to specify the attributes and 
rules, which limits its usages to only ABAC or RBAC based systems with clear 
definitions.
Third, considering the dynamic nature of access control configurations, the 
detection algorithm needs to be run every time after administrators make 
changes.

\textbf{Discussion}.
Detecting access control misconfigurations with data mining is a black-box 
approach which does not rely on the source code. This makes it possible to 
analyze the policies across multiple components of software applications and 
system environments.
However, the main limitation of conduct data mining to detect access control 
misconfigurations is the general assumption that most access configurations or logs 
are correct. 
Some parameters (e.g. support or confidence) need to be tuned to balance the 
quality and quantity of produced rules.
The other limitation is that not all the produced rules are correct, which 
incurs a burden on developers or administrators to manually identify the false 
positives.

%% file: verify.tex
\section{Verification}
\label{sec:verify}
\begin{figure}[h]
	\includegraphics[width=\linewidth]{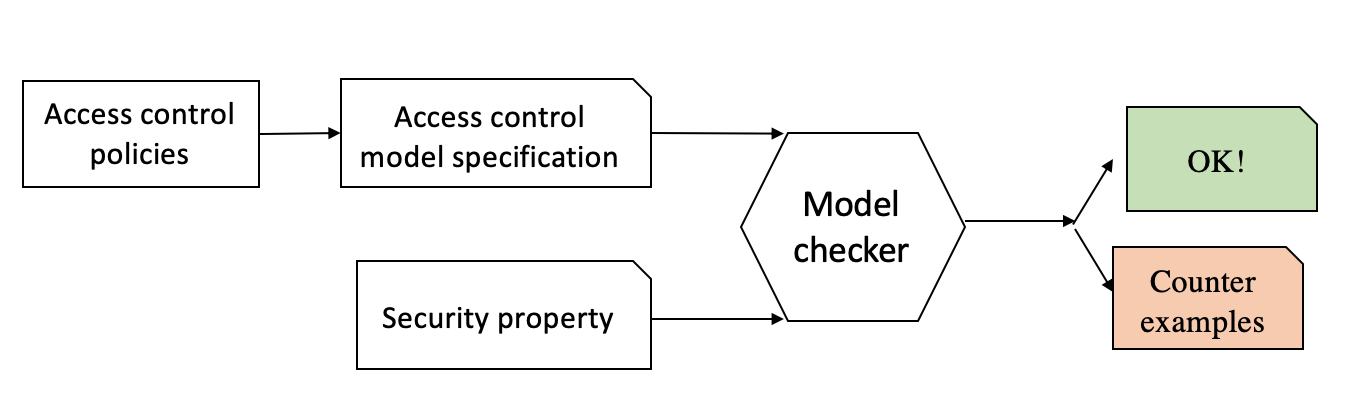}
	\caption{\textbf{Verification approach overview.} The verification approach 
	requires the sysadmins or the developers to specify the access control 
	model with a formal specification, and the security properties.
The model 
	specification can be derived from access control policies. 
	\label{fig:verify}} 
\end{figure}
Formal methods such as model checking or other verification methods were used 
to validate the correctness of access control policies. Generally speaking, the 
developers need to abstract the access control policy into specifications and 
devise security property based on the configuration goal. The verification tool 
like model checker or SAT solver will be used to find counterexamples of the 
property or proves the specification satisfies the property. Note that all the 
proposed verification techniques only apply to policies with ABAC or RBAC 
systems since specifications can describe these models well.
In this section, we discuss three representative works that analyze the 
policies with different goals and some techniques to improve the efficiency of 
verification.

\noindent\textbf{Margrave}~\cite{fisler2005verification}. 
Fisler et al. develops a framework, Margrave, to analyze the access control 
policies written in XACML for ARBAC systems. Margrave is designed for two uses. 
First, the tool can check whether 
a policy (specification) satisfies a property. Second, given two versions of 
policies, Margrave can highlight the differences between the two policies. This 
can help administrators understand the impact when making policy changes.

\begin{figure}
	\includegraphics[width=\linewidth]{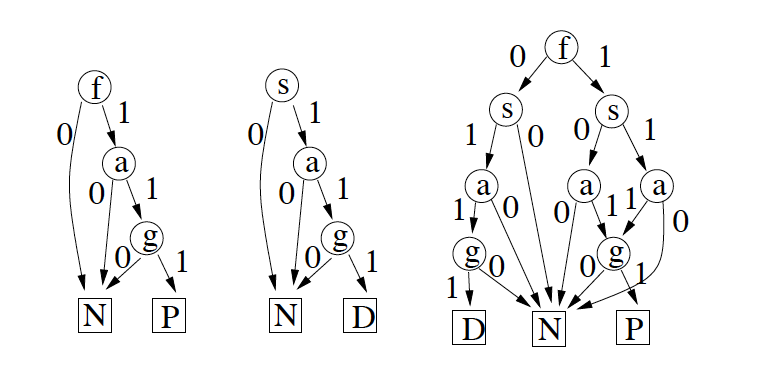}
	\caption{Sample MTBDDs in Margrave~\cite{fisler2005verification}. The left 
	two trees represent two 
	individual rules for \textbf{f}aculty and \textbf{s}tudent with 
	\textbf{g}rade and \textbf{a}ssign 
	capabilities. The tree on the right side is the combined 
	rule.\label{fig:margrave}} 
\end{figure}

To achieve these two goals, Fisler et al. designed an innovative representation 
of XACML policies, named multi-terminal binary decision diagrams (MTBDD). As 
shown in Figure~\ref{fig:margrave}. The terminal nodes represent four possible 
decisions, $\{permit, deny, not-applicable, EC\}$. $EC$ refers to environment 
constraint which can decrease the complexity by eliminating some paths in the 
MTBDD, such as ``a user can only has one role''. 
The path from the root to the 
leaf constructs a complete rule. To achieve the first goal, when the 
administrator wants to query a role with specific attributes, Margrave simply 
matches the rule from the root to the leaf to find whether the decision from 
the policy matches the desired security properties.

The second challenge is to find the changes between two versions of policies. 
To address this challenge, Fisler et al. defines a  
change-analysis decision diagram (CMTBDD), which has 16 terminals. Each 
terminal represents a state transition from the original policy's decision to 
the new decision, 
such as $permit\rightarrow permit$, $permit\rightarrow deny$, etc.
CMTBDD can be constructed by comparing the MTBDDs produced by two policies. They 
also designed several APIs to help retrieve the values of relevant attributes from 
the constructed trees.

The main limitation of Margrave is that it can only be applied to rules there 
are no dependencies, complex conditions, or requests with multiple subjects, 
even though these functions are provided by the XACML standard. These complex 
policies need complicated modeling which does not fit for simple branch 
conditions in the MTBDDs.
Besides, Margrave can only be adopted by RBAC or ABAC based systems. Developers 
or administrators need to correctly specify the property as their goal as well 
as specifications corresponding to the XACML policy. More convenient and 
easy-to-use tools may be needed to make the verification tool more practical.

\noindent\textbf{Hu et al's work}~\cite{hu2008enabling} This work is the 
representative work of applying model checking tools for security policy 
verification and generating security assurance code and test cases for testing 
purposes. The model-based verification framework requires a formal 
representation of models (e.g. roles, actions), security specifications (e.g. 
roles should or should not have access to resources in the policy to achieve 
access control goals), and security properties (e.g. high-level description of 
whether an authorization state is permitted.) 

The authors defined an access control model specification $S$ as $S=(M,F,C)$, 
where $M$ is the model containing roles and relations, $F$ is a set of features 
achieved by the access control policy, and $C$ is a set of high-level 
constraints such as $safety$ problems (i.e., $under-constraint$, too much 
access granted to certain users) or $accessibility$ problems (i.e., 
$over-constraint$, legitimate accesses are denied). 

The authors defined 
two kinds of access control model properties, access control function property 
$P_f$ describing expected function results and access control constraint 
property $P_a$ describing expected authorization states. When the security 
properties are properly defined, $S \models P_f$ means the proposed functional 
property holds on the model specification $S$, otherwise the functional 
property will be violated. For the constraint property, if $S\not\models P_a$, 
it means there is an over-constraint. In order to express the under-constraint, 
the unexpected constraint property $P_{a^-}$ needs to be designed, to show the 
specification is strictly lower than the constraint.

With the help of verification results, counterexamples can be identified for the 
property violation and used as test cases for the access control policy. This 
is called \textit{model-based testing}. Positive test cases can also be derived 
from the model checker by satisfiable examples.

The authors further implemented their method with Alloy~\cite{halpern2008using} 
model checker. The example shows that the tool can apply to RBAC standards. 
However, the developers or the sysadmins still need to translate the entities 
and relationships in the access control policies to the Alloy language. 
Specific security properties also need to be clearly defined for each kind of 
error that they are interested in. This effort is non-trivial.

Besides, Hu et al's work has several common limitations of using verification 
to detect 
incorrect access control policies~\cite{hu2008enabling}
First, it can only work on systems with RBAC/ABAC formal models which can 
easily translate the roles into the model.
Second, the model checker has the state space explosion problem when dealing 
with complex policies with many entities and relations. This can be very time-consuming 
verification even as a one-time effort. Not to mention the dynamic 
changes in access control policies every day.
The follow-up work tries to improve the efficiency with abstraction refinement~ 
\cite{jayaraman2011automatic}.

\noindent\textbf{Mohawk}~\cite{jayaraman2011automatic}. Jayaraman et al. 
proposed to apply abstraction-refinement techniques to solve the state space 
explosion problem in model checking of access control policies. The idea of 
abstraction refinement has been used in hardware verification and program 
analysis to improve efficiency, but the specific properties need to be 
defined for the abstraction and refinement steps.
\begin{figure}
	\centering 
	\includegraphics[width=\linewidth]{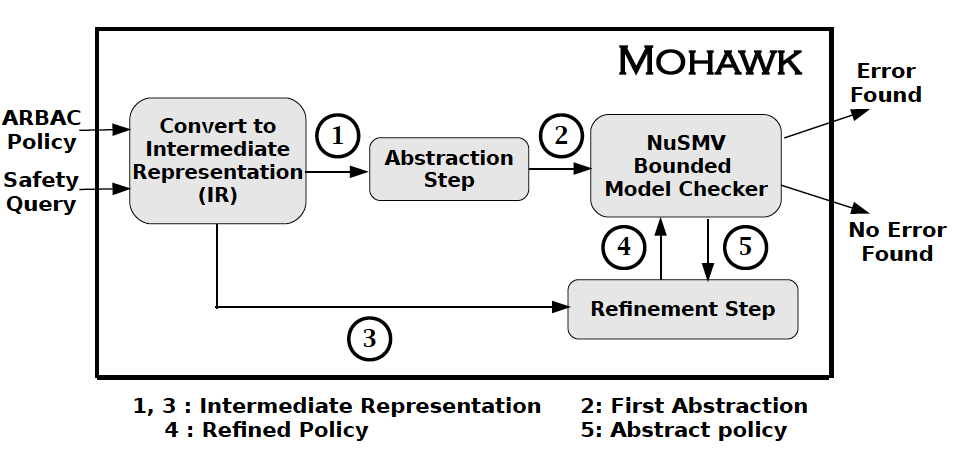}
	\caption{The architecture of 
	Mohawk~\cite{jayaraman2011automatic}.\label{fig:mahawk}} 
\end{figure}

Mohawk solves a very specific problem in access control systems - whether a 
user $u$ can become a member of role $r$ through the 
state transitions.
Consider a policy contains a set of $<\Gamma,\Phi>$, where $\gamma\in\Gamma$ is 
a state 
and $\phi\in \Phi$ 
is a state-change rule. $\gamma$ can be a state that user $u$ has a role $r$ 
that $\gamma = <u,r>$. $\phi$ can be a rule that $u$ can be assigned to another 
role by \textit{admin} based on its original role. This satisfies the requirement of 
RBAC well because all users' abilities are granted via roles. Given a policy 
and a safety query of $<u,r>$, Mohawk determines whether this is achievable and 
generates the corresponding example via abstraction and refinement steps.

Mohawk decreases the possible states to be explored, by the abstraction step. 
As shown in Figure~\ref{fig:mahawk}, Mohawk first translates the policy into a 
common intermediate representation (IR) in a special format for Mohawk. The 
orders in the IR relationships can construct a priority queue for the roles to 
be considered.
Then in the abstraction step, only the directly related and admin users, roles, 
and user assignment rules are included in the abstract policy.
This will only capture partial states and may not be able to find an error 
through the model checker. If no errors are found, the refinement step will 
refine the abstract policy, by adding relevant roles, users, and rules in the 
IR. If no more refinement can be added, the model checker will stop and 
announce there are no errors for the safety query (i.e., user $u$ can never 
become a member of $r$).

The abstraction-refinement steps can reduce the state space to be explored by 
only considering partial space related to the safety query, and adding more 
states in the refinement step to find possible errors. All the users or roles 
not relevant to the query are eliminated, which helps most especially for large 
complex policies where many entities and relationships are not related. The 
experimental results show that Mohawk scales well to large complex access 
control policies which can efficiently help administrators check the policies 
before 
rolling out to real systems.


\textbf{Discussion}.
In summary, the verification approach towards finding errors in access control 
policies can apply to RBAC or ABAC systems with well-defined roles and access 
control policies.
The advantage of this method is that if the verification finds no violations to 
the security property, the security property will be assured.
However, the verification approach has several disadvantages in terms of 
error-finding scope, human efforts, time and space efficiency, and scalability 
to real systems.
First, the verification requires an abstract model representation and security 
property specifications. This makes it hard to apply in systems other than 
ABAC or RBAC systems where no formal models can describe them well.
Second, the verification only finds inconsistencies based on the security 
property, but misconfigurations may still exist even though the policy is 
consistent. For example, as described in \cite{das2010baaz}, some users may be 
mistakenly added certain abilities by ad-hoc changes, which is hard to be 
modeled or specified in a security property.
Third, access control goals (security properties) may also change over time. 
This means each time the access control policy changes, the specifications and 
the security property also need to be changed. Fourth, even though some 
techniques have been developed such as abstraction refinement described in 
Mohawk\cite{jayaraman2011automatic}, it still takes a large amount of time to 
comprehensively verify the policies.

%% file: testing.tex
\section{Testing}
\label{sec:test}
Access control policies may have errors, just like programs may have bugs. 
Software testing is a good analogy for policy testing as shown in 
Figure~\ref{fig:testing}.
These test cases can detect
access control policy errors before
administrators rolling out their changes. We will discuss two kinds of testing 
approaches based on the source of test oracles. The test oracles can be 
specified by formal models or from human input.

\subsection{Model-based Testing}
The model-based testing approach generates test cases based on the formal 
models. As discussed in Section \ref{sec:verify}, counter examples will be 
found if the model does not satisfy the property. The model specification and 
the safety property can check whether the response is expected or not.The key 
idea is to use the generated counter examples as the test cases.

Two kinds of test cases can be generated. The first is negative test cases, 
which represent test cases which do not satisfy the security property.
The 
second is the positive test cases which satisfy the security property.
To generate the positive test cases, we can negate the original desired 
security property, then the generated counter examples will represent the 
desired access control results.

\subsection{Human-assisted Testing}
This line of work requires human experts to examine whether the test responses 
are expected or not.
Previous work has made efforts towards automatically generating high-quality 
test cases to capture the faults in the policies. 
Most work has focused on access control 
policies written in a formal standard like XACML with the help of mutation 
testing. We describe two lines of work from (1) automatically 
generating requests~\cite{martin2007automated, martin2007fault} to (2) 
strategies to improve the effectiveness of test 
cases~\cite{bertolino2012automatic, bertolino2014coverage}.

\begin{figure}
	\includegraphics[width=\linewidth]{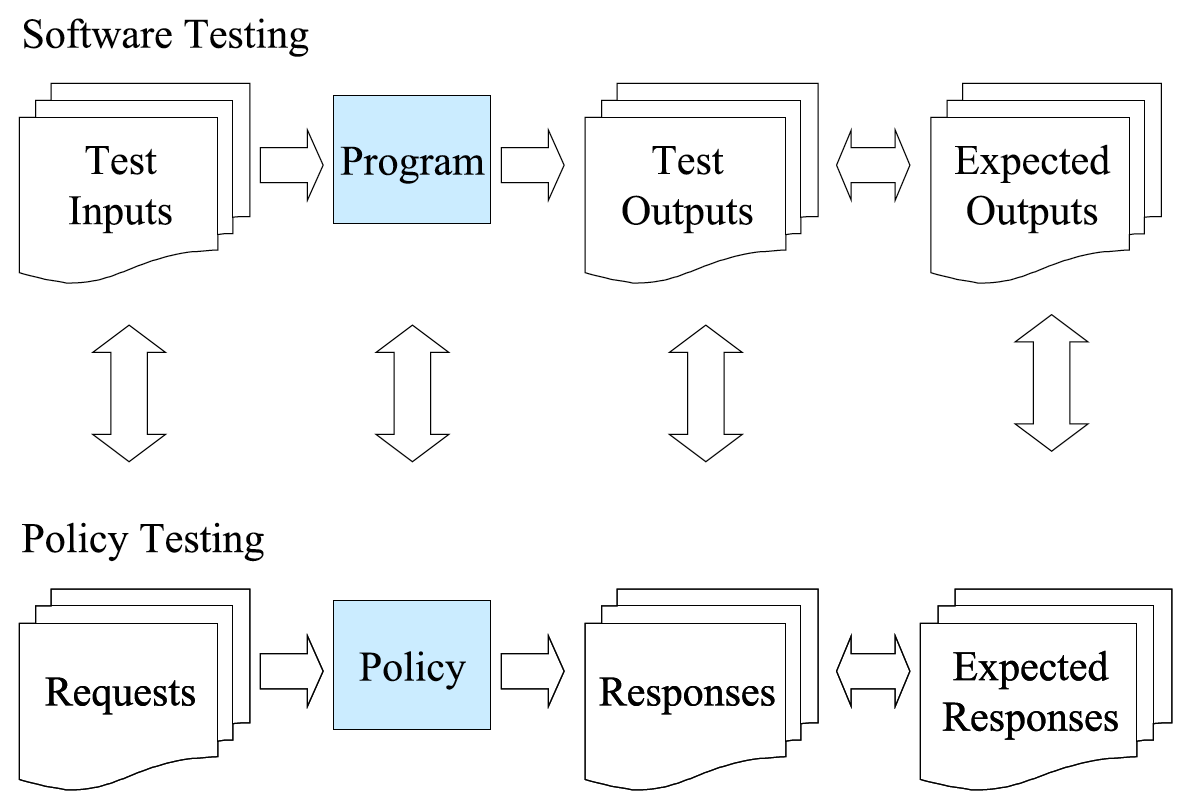}
	\caption{Workflow of policy testing~\cite{martin2007fault}. Policy testing 
		is similar to software testing. The test cases become requests. 
		Misconfigurations can be detected by comparing the responses with 
		expected responses
		\label{fig:testing}} 
\end{figure}

\noindent\textbf{Martin et al's work}\cite{martin2007automated, 
martin2007fault}. Martin et al. are the first to propose testing access control 
policies via the mutation testing approach. They found that, similar to software 
testing which aims to find faults in programs which leads to abnormal behavior, 
access control policies can also be tested. 
A fault in access control policy 
will lead to wrong access control decisions, which may lead to unauthorized 
sensitive data access. Their works make the first step to test access control 
policies via mutation testing~\cite{jia2010analysis}.

The test cases for policy testing are 
requests and the generated responses (i.e., Permit or Deny) will be compared 
against expected responses, as shown in Figure~\ref{fig:testing}. 
In order to evaluate the quality of generated test cases, they first defined 
several metrics, including \textit{policy coverage},\textit{ rule coverage}, 
\textit{condition coverage}, 
\textit{mutant killing ratio}. The first three metrics are rather intuitive. 
The mutant killing ratio is from the mutation testing~\cite{jia2010analysis} 
for software testing. 
A \textit{mutant} of a policy is a new version by modifying small 
parts of the policy. A mutant kills the original policy if the response 
evaluated on the new policy differs from the response evaluated on the original 
policy. A higher mutant killing ratio usually means a better quality of 
generated test cases. 
The intuition is that if a test case can not differentiate mutants and the 
original policy, it is less likely to detect the fault.

\begin{table}
	\begin{tabular}{l l}
		
		\toprule
		Metrics & Definition\\
		\midrule
		Policy coverage & The percentage of policies involved \\
		& in evaluating generated requests.\\
		Rule coverage & The percentage of rules involved \\
		& in evaluating generated requests.\\
		
		Condition coverage & The percentage of conditions involved \\
		& in evaluating generated requests.\\
		
		Mutant killing & The percentage of mutants killed \\
		ratio & by generated requests.\\
		\bottomrule
	\end{tabular}
	\caption{Coverage metrics for request generation in policy testing.}
\end{table}

In~\cite{martin2007automated}, Martin and Xie designed a tool named Crig 
towards automatically generating requests (test cases) via change-impact 
analysis~\cite{fisler2005verification}. This is divided into four steps:
\begin{itemize}
	\item First, Crig synthesizes different versions of the original policy. 
	The new versions can be generated by (1) negating the decision for each 
	rules inside the 
	policy one by one, or (2) only preserving one rule at a time and negating 
	it. Both approaches can achieve high rule coverage.
	\item Second, Crig takes the generated new versions of policy and applies the 
	change-impact analysis tool on it. This is done with the help of 
	Margrave~\cite{fisler2005verification}. By feeding the different policies 
	into Margrave, it will generate requests which will have different 
	decisions in the two policies.
	\item Third, with the requests generated by Margrave, Crig rebuilds the 
	concrete requests by using solvers if the request is only at the abstract level.
	\item Fourth, Crig will reduce the number of requests by finding a minimal 
	set of requests which will still met the desired coverage requirement, such as 
	structural coverage requirement.
\end{itemize}

Note that the generated requests (test cases) still need to be manually 
inspected by the administrators because the requests are just generated based 
on the extended behavior of the current policy. If the policy itself contains 
errors, the generated test cases will prevent correct policy usages.

Martin et al. further formalize the above testing idea by proposing a generic 
fault 
model of access control policies~\cite{martin2007fault}. The proposed fault 
model divides the faults at two levels: syntactic errors and semantic errors. 
Syntactic errors refer to the typos in attribute values. These errors will 
impact the value of conditions or target elements (subject/action/resources). 
Semantic errors are the errors related to incorrect use of logical operators, 
such as the rule combining algorithm (e.g. first-applicable) or rule effects 
(Permit or Deny). 
They further proposed mutation operators based on specific error types. To 
reduce the number of mutants for better efficiency, they used the change-impact 
analysis tool~\cite{fisler2005verification} to compare two policies whether they are 
the same, i.e., having 
the same responses for all inputs (requests).

\noindent\textbf{Bertolino et al.'s work~\cite{bertolino2012automatic, 
bertolino2014coverage}} Bertolino et al. put additional efforts on improving 
the effectiveness and decreasing the total number of automatically generated 
test suite, by allowing customize coverage target~\cite{bertolino2012automatic} 
and defining smarter coverage criterion~\cite{bertolino2014coverage}. 

Bertolino et al. proposed two new methodologies for generating requests with 
the idea of combinatorial testing~\cite{bertolino2014coverage}. The first 
method is named \textit{Incremental XML Partition Testing~(XPT)}. The main idea 
is to (1) limit the possible values with only one 
\texttt{<AttributeValue>} and (2) 
eliminate the impossible combinations of schema based on the constraints in the 
XACML functions that no rules will be matched.
The second method is named \textit{Simple Combinatorial Testing}. This method 
also requires the possible values of subjects, actions, resources and 
environment. The maximum number of requests generated by Simple Combinatorial 
Testing is the cardinality of the four-wise combination set of four 
elements. Users can customize the values of interest in the set of four 
elements to decrease the number of generated test cases. 

However, the disadvantages of combinatorial testing are also obvious. First, it 
requires analysis to capture the possible values of the elements, which relies 
on the deep domain knowledge of the tester. Some automated tools may be helpful 
in constructing the initial sets, but they still need manual inspections for 
high-quality assurance. Second, the exploration space will 
be large in real policies considering the possible values for each element.

Bertolino et al. further proposed a smart coverage criterion 
in~\cite{bertolino2014coverage} for better test case selection. 
The main insights are that instead of focusing on coverage of policies, rules 
or conditions, the \textit{targets} inside the rules, policies or policy 
sets are more important because if the request does no match to a target in the 
policy, it will not be evaluated. 
With the new coverage criterion, they proposed test cases selection algorithm 
to reduce the total number of test cases as well as preserve a high fault detection 
rate (as indicated by the mutant killing ratio in the experiments).

In summary, these two works ~\cite{bertolino2012automatic, 
bertolino2014coverage} reduces the total number of test cases by allowing 
customization of possible value sets in combinatorial testing, or using smart 
coverage criterion to select relevant requests while preserving optimal fault 
detection effectiveness at the same cardinality level.

\textbf{Discussion}.
The testing approach mainly finds the faults in access control policies with the 
help of generated test cases. 
However, it has several limitations.
First, the generated requests still need to be manually verified by the 
administrators. Even though efforts have been made to increase the quality and 
decrease the number of test cases, it still takes huge efforts to verify the 
correctness of policies.
Second, access control policies will be changed due to the dynamic needs of the 
systems. This means that previous test cases may not be correct after the 
policy has been changed. The test cases may be updated every time or 
re-generated and re-verified everytime the administrator makes changes, which is 
contradicting the goal of automatic test cases generation.
Third, automatically generating test cases requires a formal standard like 
XACML, which is not widely used. 
It becomes a huge obstacle for the administrators to switch to ABAC or RBAC 
with a formal definition of roles, actions, etc.
All these issues make the testing approach less attractive for practitioners. 
Nonetheless, the comprehensive tests may still be needed for static access 
control policy environment, or critical missions like aerospace engineering.

%% file: ccs.tex
\section{Our work}
\label{sec:ccs}

We present P-DIFF, a tool to inferring access control configuration changes and 
behavior changes from access logs~\cite{xiang2019towards}. As discussed in 
previous sections, current detection techniques have several limitations 
that most tools (1) can only detect inconsistencies of access control policies, 
(2) only work on one system with the same format of configurations, and (3) 
needs additional efforts from developers and administrators to specify 
configuration format or even comprehensive security goals. These limitations 
make them hard to be used in real-world systems, where the access control 
policies are dynamically changing and scattered into different components.
\begin{figure}[h]
	\includegraphics[width=\linewidth]{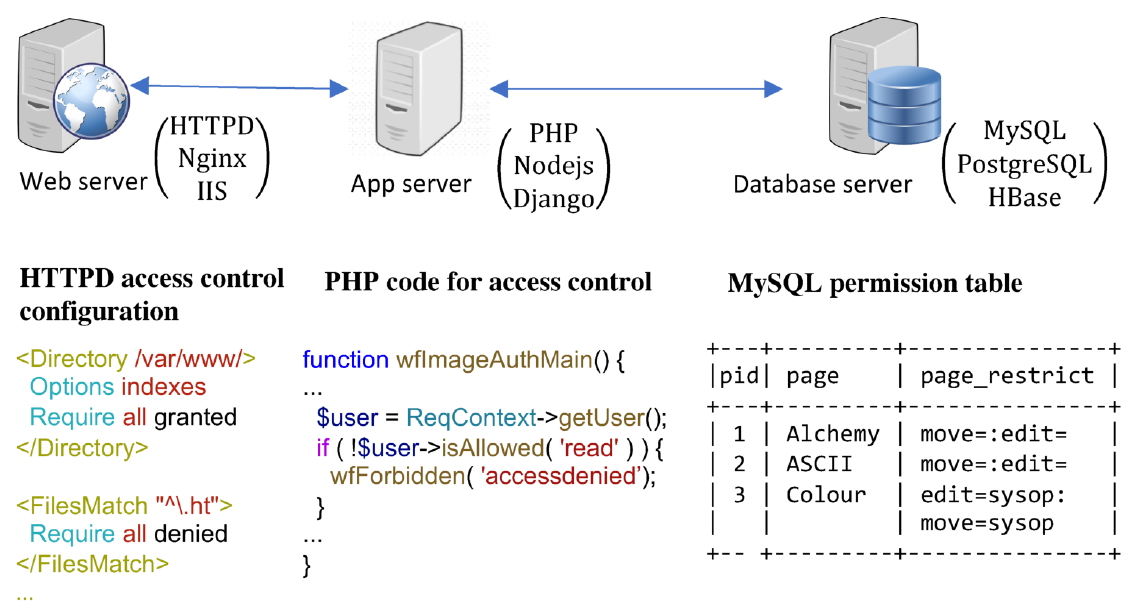}
	\caption{Heterogeneous access control configurations and code 
		implementations in real-world web server 
		scenario~\cite{xiang2019towards}. Each component has its own security 
		policies, which makes it more challenging to come up with a uniform 
		representation.\label{fig:pdiff1}} 
\end{figure}

\begin{figure}[h]
	\includegraphics[width=\linewidth]{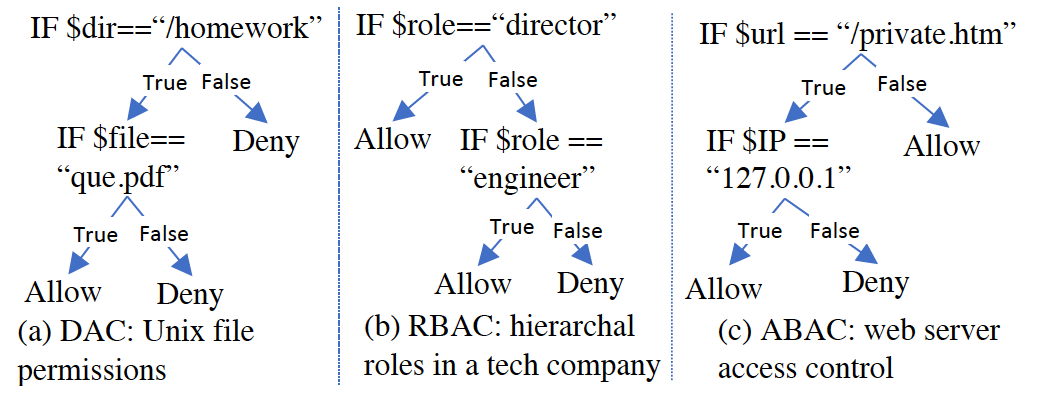}
	\caption{Decision tree can represent the intricate rules with attributes 
	like URL, file path and roles in all three access control models
~\cite{xiang2019towards}.\label{fig:pdiff2}} 
\end{figure}

P-DIFF aims to infer access control policies to monitor policy changes for two 
reasons. First, the access log contains enough information to be able to infer 
correct policies. Attributes such as user, group, IP addresses all exist in the 
logs. The access result represented in the access logs is the 
\textit{end-to-end} access 
control result of the whole system.
Second, it is hard and impractical to come up with a unified 
representation for access control configurations across multiple components. As 
shown in Figure~\ref{fig:pdiff1}, a real-world website involving webserver, 
appserver and database has complex access configurations in each component, 
resulting in the desired security goals. Misconfigurations in either component
can result in disastrous consequences.

\begin{figure}
	\includegraphics[width=\linewidth]{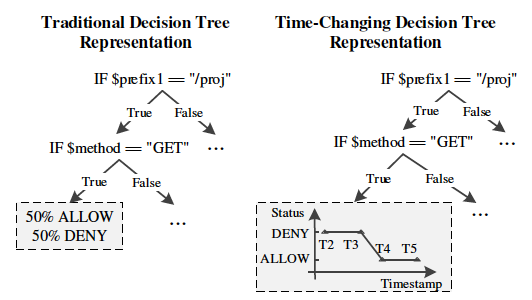}
	\caption{Decision tree can represent the intricate rules with attributes 
		like URL, file path and roles in all three access control models
		~\cite{xiang2019towards}.\label{fig:pdiff3}} 
\end{figure}

P-DIFF addresses the heterogeneous challenge by inferring rules from the access 
control logs using a novel decision tree algorithm. The decision tree is 
suitable for representing access control goals because of the hierarchical 
nature 
of trees. As shown in Figure \ref{fig:pdiff2}, access control models in three 
different applications, can all be represented by a decision tree structure. 
This also makes the inferred rules explainable for the system administrators to 
be used for validation. The terminal leaf node encodes the decisions of Allow 
or Deny. Similar to \cite{shaikh2010inconsistency, fisler2005verification}, the 
paths from the root to the leaf node represents an inferred rules or policy. 
Compared to data mining algorithms such as associate rule learning used 
in~\cite{bauer2011detecting,zhang2014encore}, the decision tree can capture the 
decision process in a hierarchical manner.

Considering the dynamic nature of access control, sysadmins need to make 
changes to the access control policy to satisfy needs like user or resource 
access changes. 
P-DIFF can help with the behavior validation by encoding time information in 
the decision tree.
We propose a time-changing decision tree (TCDT) where each node contains the 
whole access history which matches the specific rules from the root to the 
node. As shown in Figure~\ref{fig:pdiff3}, traditional decision tree uses the 
percentage of Allow/Deny in each node to represent the decision information. 
All the information related interesting changes are lost with only 
representation of percentages. 
TCDT records the history information by associate the decision with timestamps, 
such that useful information related to changes (e.g., Deny to Allow). Then the 
history changes for individual rules are captured. 
Correspondingly, 
instead of optimizing entropy or information gains, TCDT split a node into two 
subsets based on the number of changes, with the intuition that the total 
number of changes in the rules are as least as possible. For example, when all 
file access decisions under a directory changes, it is more likely to be a 
directory level permission change, instead of for each files. This will 
optimize the total number of rules found in the TCDT.

Despite the advantages, P-DIFF has several limitations. First, P-DIFF
infers the access control policies from access logs, which limits its scope to 
the information contained in the logs. If the logs miss important information 
like users, actions, P-DIFF can only infer partial rules which may not be 
consistent with the actual policies.
Second, P-DIFF generates policies with access logs in an end-to-end fashion. 
Therefore, the inferred policy does not necessarily represent the original 
access control policy. The inferred policy is the resulting states 
from access control policies in multiple components. For example, a user may 
have different definitions and access control policies in PHP code and 
databases, however, the logs are obtained from webserver with the users in the 
webserver configuration. Even though the rules inferred by TCDT represents the 
access control policy resulting from the whole system, sysadmin may face 
difficulties in using the inferred policy to find the possible 
misconfigurations. 
Third, P-DIFF only detect changes after there is an access causing rule 
changes, otherwise, the inferred rule will not change. Sysadmins will be 
notified once there exists such rule changes and need to take measures as soon 
as possible to defend against future damages.

%% file: discussion.tex
\begin{table*}[!htbp]
\begin{adjustbox}{width=1\textwidth}

	\begin{tabular}{l l l l l l l}
		
		\toprule
		
		Techniques & Studied work & Input  & Human Efforts & Static/Dynamic
		&Security/Accessibility& Main Techniques\\
		\midrule
		\multirow{5}{*}{Data Mining}&Bauer's\cite{bauer2011detecting}&Access 
		logs & Low  & Static& Accessibility& Associate Rule Mining\\
	 	&Bazz~\cite{das2010baaz} & User-Group 
	 	Mapping & Medium& Static &Both &Group mapping/clustering\\
		&EnCore\cite{zhang2014encore} & AC configurations& Medium& Static&Both 
		& Associate Rule Mining\\
		&Shaikh's\cite{shaikh2010inconsistency}& AC policies& 
		Low&Static&Both&Decision Tree\\ 
		&P-DIFF\cite{xiang2019towards} & Access	logs & Low&Both& Both& 
		Time-Changing Decision 
		Tree\\
		\midrule
		
		\multirow{3}{*}{Verification}&Margrav~\cite{fisler2005verification}& 
		AC models, security properties& High& Both& Both & Decision diagrams 
		MTBDD\\
		&Mohawk\cite{jayaraman2011automatic}& AC models, safety 
		query& High& Static& Both& Model Checking\\\
		&Hu's\cite{hu2008enabling} & AC models, security 
		properties& High & Static&Both& Model Checking\\
		\midrule
		\multirow{2}{*}{Testing}&Martin's\cite{martin2007automated,
			martin2007fault} & AC policies& High&Static&Both& Mutation Testing\\
		&Bertolino's\cite{bertolino2012automatic,                       
		bertolino2014coverage}& AC policies& High&Static&Both& Combinatorial 
		testing\\
		\bottomrule
	\end{tabular}
\end{adjustbox}

	\caption{\textbf{Summary of the work studied in this survey}. Input is the 
	prerequisites of this approach, usually needs to be specified by developers 
	or administrators. The level of human efforts involves two aspects: (1) 
	obtaining relevant data to use the approach; (2) validation of the produced 
	results. Static or dynamic refers to whether the techniques can be applied 
	on access control configuration with changes all the time. Security 
	misconfigurations refer to the errors in configurations causing too much 
	access granted to users, while accessibility misconfigurations cause 
	legitimate accesses being denied.
	\label{tab:sum}}

\end{table*}

\section{Summary and Future Work}
\label{sec:discuss}
In this section, we first summarize the surveyed works covering detection
techniques of data mining, verification and testing. Based on the limitations 
of current works, we further identify
several directions that are valuable for future research.

\subsection{Summary of Previous work}

We summarize the previous works in Table \ref{tab:sum}. This survey covers 
three approaches detecting access control misconfigurations. The three 
approaches obtain the representation of the ground truths of security goals 
with different focuses, therefore require different inputs from developers or 
administrators.

The data mining approach infers correct security goals from mostly correct 
configurations or access logs. This approach usually requires less human effort 
because no ground truth needs to be specified by humans. However, developers or 
administrators still need to provide some format to parse the log or 
configurations. However, the inferred ground truth may be stale or incorrect. 
Therefore, the detected misconfigurations still need to be verified by the 
sysadmins. The false positives of the detected misconfiguration may cause a 
waste 
of time and cause frustration.

The verification approach requires humans to specify the representations of 
security goals as well as the security properties, which needs high level of 
human efforts. Most verification techniques can only focus on static 
verification which means the security specifications and properties may be 
changed every time when the access control policy changes.  
Margrav~\cite{fisler2005verification} improves this by comparing the changes 
between two versions of policies via change-impact analysis which partially 
mitigates this issue. 

The testing approach assumes the initial access control policy is correct and 
generates test cases based on the correct policy. These approaches try to 
automatically generate test cases with good coverage of security policies. 
However, the test cases need to be manually verified by the sysadmins to be of 
real use. Therefore, Bertolino et al's\cite{bertolino2012automatic,     
bertolino2014coverage} works focus on optimizing the total number of test cases 
without decreasing the test cases quality too much. However, the main 
limitation is that the test cases works for previous access control policy may 
be wrong since the security goals change all the time based on the system's 
needs, thus test cases need to be re-generated every time the security goal 
changes.

\subsection{Future Directions}

We present two possible future directions to be explored based on the usability 
of configurations and the prevalent cross-component access control errors.

\subsubsection{Usable Configurations}
The complexity of software applications is increasing and the access control 
configurations become more and more intricate. 
Many companies have system 
administrators responsible for maintaining and updating the systems, as well as 
fixing issues. However, many misconfigurations are introduced when they are 
fixing issues~\cite{xu2017system}. This suggests improving the usability of 
configurations in both design and diagnosis. 

Many studies~
\cite{felt2011android,wijesekera2015android,shen2021can} 
have focused on improving the usability of access control on mobile platforms 
(e.g., Android, iOS), but only a few works
~\cite{xiang2021detecting,shen2022automatic,shenmultiview,shenimproving} have focused on 
improving the usability of access control configurations for sysadmins.
Some previous works have focused 
on improving the design of configurations so that sysadmins are less likely to 
make mistakes~\cite{xu2015hey, keller2008conferr}, or helping them diagnose the 
configuration problem with better tools~\cite{wang2004automatic, 
rabkin2011precomputing}. 
Future research may look into how to generate better diagnostic information 
or visualize the security impact of configuration change in real systems, which 
can help prevent the misconfigurations in the first place.

\subsubsection{Cross-component misconfigurations}
Previous works mainly detect the misconfigurations in one system written in one 
language. However, due to the complexity of today's software systems, the 
access control configurations are scattered into multiple components in various 
formats. Therefore, it is hard to model the configurations in multiple 
components with a unified representation. Even though some current 
works~\cite{xiang2019towards} focus on the end-to-end monitoring of access 
control results, systems still need to look into multiple components to 
understand how the access result happens. Future works can look into how to 
jointly analyze the configurations from multiple components and pinpoint the 
misconfigurations in the right component.